\documentclass[preprint,proceedings]{rmaa}

\usepackage{paralist}

\usepackage{psfrag,color}

\SetYear{2003}
\SetConfTitle{IAUC 194: Compact Binaries in the Galaxy and Beyond}

\title{A ULX in NGC\,4559: a ``mini-cartwheel'' scenario?} 

\author{
  R. Soria,\altaffilmark{1}
  M. S. Cropper,\altaffilmark{1}
  and M. W. Pakull\altaffilmark{2}}

\altaffiltext{1}{Mullard Space Science Laboratory, University College London}
\altaffiltext{2}{Observatoire de Strasbourg, France.}

\shortauthor{R. Soria, M. Cropper, \& M. Pakull}
\shorttitle{ULX in NGC\,4559}

\fulladdresses{
\item Roberto Soria, Mullard Space Science Laboratory,
  University College London, Holmbury St Mary, Surrey, RH5 6NT, UK
  (\email{Roberto.Soria@mssl.ucl.ac.uk}).
\item Mark Cropper, Mullard Space Science Laboratory,
  University College London, Holmbury St Mary, Surrey, RH5 6NT, UK
  (\email{msc@mssl.ucl.ac.uk}).
\item Manfred Pakull, Observatoire de Strasbourg,
  11, rue de l'Universit\'{e}, F-67000 Strasbourg, France 
  (\email{pakull@astro.u-strasbg.fr}).
}

\listofauthors{R. Soria}
\indexauthor{Soria, R.}
\indexauthor{Cropper, M. S.}
\indexauthor{Pakull, M. W.}

\abstract{We have studied an ultraluminous X-ray source 
(ULX) in NGC\,4559 with {\it XMM-Newton}, and its peculiar 
star-forming environment with HST/WFPC2.
The X-ray source is one of the brightest in its class 
($L_{\rm x} \approx 2 \times 10^{40}$ erg s$^{-1}$). 
Luminosity and timing arguments suggest a mass $\ga 50 M_{\odot}$ 
for the accreting black hole. The ULX is located 
near the rim of a young (age $< 30$ Myr), 
large (diameter $\approx 700$ pc) ring-like star forming complex 
possibly triggered by the impact of a dwarf satellite 
galaxy through the gas-rich outer disk of NGC\,4559.
We speculate that galaxy interactions (including 
the infall of high-velocity clouds and satellites 
on a galactic disk) and low-metallicity environments 
offer favourable conditions for the formation 
of compact remnants more massive than ``standard'' 
X-ray binaries, and accreting from a massive Roche-lobe filling 
companion.}

\resumen{}

\addkeyword{Galaxies: spiral}
\addkeyword{Galaxies: Individual: Name: NGC\,4559}
\addkeyword{Galaxies: Interactions}
\addkeyword{X-Rays: Galaxies}

\begin{document}
\maketitle

\section{A ``canonical'' ULX in NGC\,4559}
\label{sec:intro}

Located at a distance of $\approx 10$ Mpc (Sanders 2003), 
the late-type spiral NGC\,4559 hosts 
two ultra-luminous X-ray sources (ULXs) 
with isotropic luminosities $\ga 10^{40}$ erg s$^{-1}$ 
(Vogler, Pietsch \& Bertoldi 1997).
Its low foreground Galactic absorption 
($n_{\rm H} \approx 1.5 \times 10^{20}$ cm$^{-2}$; Dickey \& Lockman 1990)
makes it a suitable target for X-ray and optical studies 
aimed at determining the nature of these sources, 
and their relation with the host environment.
A study of the X-ray timing and spectral properties of the two 
brightest X-ray sources in NGC\,4559, based on {\it {\it XMM-Newton}} 
and {\it Chandra} data, is presented in Cropper et al.~(2004). 
We focus here on the brighter of the two 
sources, X7 (using the naming convention of Vogler et al.~1997), 
which presents some of the ``canonical'' features 
of ULXs in star-forming galaxies. It has an isotropic luminosity 
$L_x \approx 2 \times 10^{40}$ erg s$^{-1}$ in the 0.3--10 keV band, 
suggesting a mass $\sim 10^2 M_{\odot}$ for the accreting object.
Its X-ray spectrum is well modelled by a power-law (photon 
index $\Gamma = 2.2$) with a ``soft-excess'' below 0.7 keV.
If the soft component is modelled as a simple 
blackbody or multicolor blackbody (standard disk spectrum), 
we obtain a characteristic temperature $kT_{bb} \approx 0.12$ keV, 
similar to what is found in some other bright ULXs.
Furthermore, we detect a feature at $\approx 30$ mHz in its 
power-density spectrum, which may be another indication 
of a high mass, $\ga 50 M_{\odot}$ (Cropper et al.~2004).

\begin{figure}[!t]
  \includegraphics[width=\columnwidth]{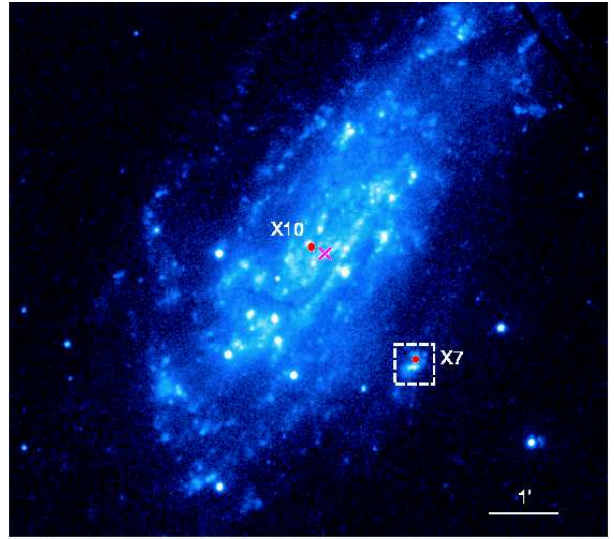}
  \caption{Both X7 and X10 in the spiral galaxy NGC\,4559 
have luminosities $\ga 10^{40}$ erg s$^{-1}$
In particular, X7 is located in a peculiar star-forming complex 
at the outer edge of the disk, as shown in this UV image 
from the Optical Monitor onboard {\it XMM-Newton}. 
The star-forming region around X7 is shown in greater detail in Fig.~2.
Here, and in the following images, North is up, East is to the left. 
At the distance of NGC\,4559, 
1\arcsec\ $\approx 45$ pc.}
  \label{fig:simple}
\end{figure}

\begin{figure}[!t]
  \includegraphics[width=\columnwidth]{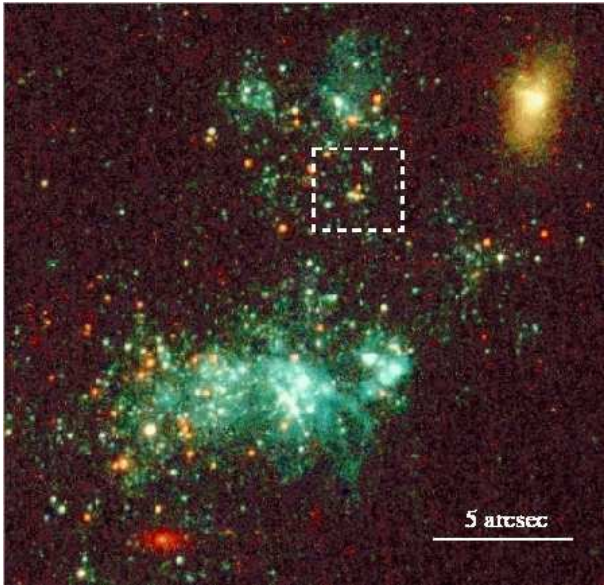}
  \caption{HST/WFPC2 true-color 
image of the star-forming complex near the X7 ULX. Filters are: 
f450w (blue); f555w (green); f814w (red). 
A likely dIrr satellite galaxy is visible near the top right corner.
The dashed rectangular region is shown in greater detail in Fig.~3.}
  \label{fig:simple}
\end{figure}

\begin{figure}[!t]
  \includegraphics[width=\columnwidth]{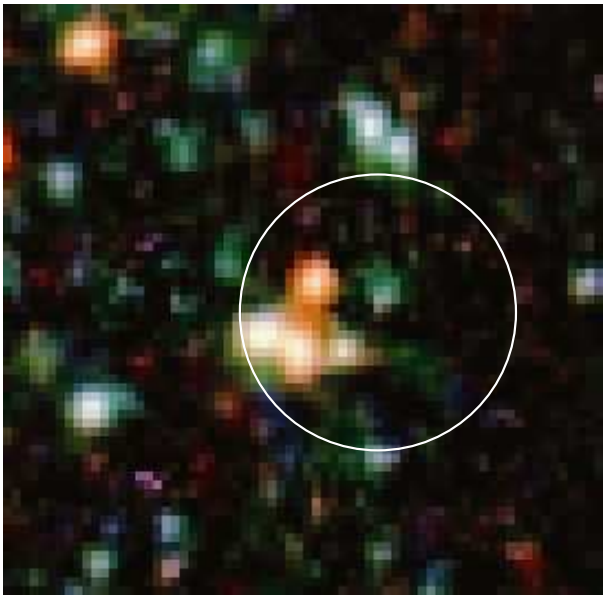}
  \caption{Zoomed-in HST/WFPC2 image showing the possible 
optical counterparts of X7 (Soria et al. 2004).
The {\it Chandra} error circle for the X-ray source has 
a radius of 0\farcs7.}
  \label{fig:simple}
\end{figure}

\section{A peculiar star-forming complex near the X7 ULX}
\label{sec:EPS}

X7 is located in a large (diameter $\approx 700$ pc), 
isolated star-forming complex,  
at the outer edge of the stellar disk of NGC\,4559 (Fig.~1).
From HST/WFPC2 observations, we infer (Soria et al.~2004) 
an age distribution between 5 and 30 Myr 
for the local stellar population, dominated by OB main sequence 
stars and red supergiants (Figs.~2-3). 
Both the optical color distribution 
and the low blue-to-red supergiant ratio (Soria et al.~2004)
are consistent with low (SMC-type) metal abundance 
(Langer \& Maeder 1995). This is in agreement with 
the low metal abundance inferred from our X-ray study 
(Cropper et al.~2004).
A CHFT H$\alpha$ observation (Fig.~4) shows more clearly 
the structure of this HII complex. The shell-like structure suggests 
that an expanding wave of star formation has recently moved 
from a centre (which appears to be near but not coincident 
with the ULX) outwards. 
Continuous star formation at a rate of $\sim 10^{-2}$ 
$M_{\odot}$/yr over the last 30 Myr would account 
for the integrated luminosity and the observed number 
of O stars (from Starburst99 models, Leitherer et al. 1999).
We estimate a mass in stars of a few$\times 10^5$--$10^6 
M_{\odot}$, depending on the assumed IMF.
The total mass (stars plus swept-up gas) 
is probably an order of magnitude higher.

Was the ULX somehow responsible for triggering 
this peculiar star-forming complex, or are they both 
consequences of another external factor? 
Large, isolated shell- or ring-like 
star-forming complexes of comparable size ($500$--$1000$ pc) 
and age ($10$--$30$ Myr) have been found in other nearby 
spiral galaxies: for example in NGC 6946 (eg, Larsen et al. 2002), 
and, on a smaller scale, in M83 (Comer\'{o}n 2001). Gould's Belt 
in the Milky Way is also similar.
From our preliminary analysis of the HST data, 
we estimate an integrated magnitude $M_B \approx -13.5$ 
for the star-forming complex in NGC\,4559, a factor of two 
brighter than Gould's Belt and a factor of four fainter 
than the NGC\,6946 complex. The latter has a young super-star 
cluster at its centre, while Gould's Belt and 
the complex in M\,83 contain only OB associations. 
None of them has a ULX. This suggests that the ULX 
is not the cause or the driving force 
of the star-forming complex in NGC\,4559.

Possible explanations for the initial triggering of such star-forming 
complexes are (Elmegreen, Efremov \& Larsen 2000; 
Larsen et al.~2002): the collapse of a "supergiant molecular cloud" 
at the end of a spiral arm; a hypernova explosion (which  
in our case might also have been the progenitor of NGC\,4559 X7); 
or the infall of a high-velocity HI cloud or satellite galaxy through 
the outer galactic disk. In all cases, the initial perturbation 
creates a radially expanding density wave or ionization front, 
which sweeps up neutral interstellar medium. Clustered 
star formation along the expanding bubble rim is triggered by 
the gravitational collapse of the swept-up material 
(eg, Elmegreen \& Lada 1977; Whitworth et al.~1994).

The large size of the complex in NGC\,4559, its location 
in the outer disk, the lack of other star-forming regions 
nearby, and the absence of diffuse 
X-ray emitting gas inside the star-forming complex seem 
to favour the collision hypothesis over the hypernova model 
(Tenorio-Tagle et al.~1986, 1987).
The impact of an $\approx 3 \times 10^5 M_{\odot}$ HI cloud 
on the Milky Way disk was simulated (Comer\'{o}n \& Torra 1994) 
to explain the formation of Gould's Belt. They show 
that, after $\approx 30$ Myr, the mass of the cold swept-up 
material is comparable or larger than the mass of the impacting 
cloud.

\begin{figure}[!t]
  \includegraphics[width=\columnwidth]{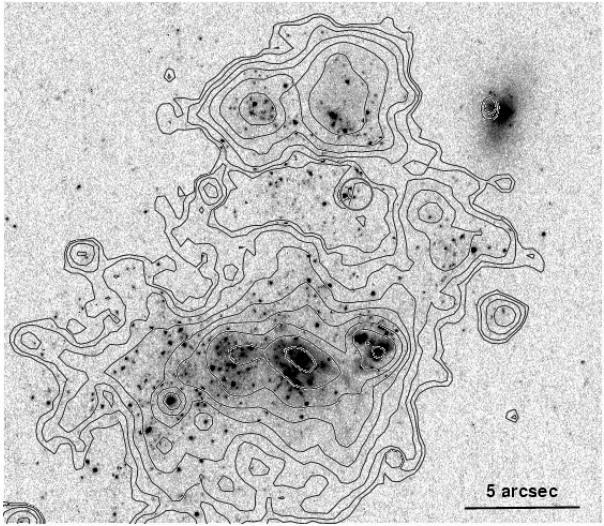}
  \caption{The ULX is located inside, although not at the center, 
of a large ring-like HII region. Greyscale image: HST/WFPC2 (PC chip), 
f450w filter ($\approx B$). Contours: H$\alpha$ image from the 
4-m CHFT. We speculate that the the dIrr galaxy seen 
at the north-west corner 
of the field is responsible for triggering this expanding 
wave of star formation, as it plunged through the gas-rich 
outer disk of NGC\,4559 some 30 Myr ago.}
  \label{fig:simple}
\end{figure}

\section{A dwarf galaxy plunging through the disk?}
\label{sec:cartwheel}

Perhaps the most intriguing result of our optical study 
of the X7 environment is that we do indeed see an object 
that could have plunged through the gas-rich disk of NGC\,4559, 
triggering the expanding density wave.
The culprit could be the yellow galaxy located $\approx 7$\arcsec\  
north-west of the ULX (Fig.~3).
This object cannot be a large background elliptical, because its isophotes 
are too irregular. Its shape, size and luminosity are consistent 
with a dwarf irregular (or, possibly, a tidally-disturbed 
dwarf elliptical) at approximately the same distance 
as NGC\,4559. We cannot presently rule out the possibility 
that it is a chance line-of-sight coincidence, but 
we suggest that the most natural interpretation  
is a small satellite galaxy of NGC\,4559.
Existing HI radio observations (WHISP survey) 
do not show any large-scale velocity distortions, 
but the satellite dwarf is perhaps too small to influence 
the galactic kinematics significantly.
We are planning optical spectroscopic observations 
to determine the kinematics and distance of the dwarf galaxy.

If the dIrr galaxy is indeed physically associated 
with the star-forming complex, its integrated luminosity is 
$M_B \approx -10.7$, with color indices 
$B-V \approx 0.47$ and $V-I \approx 0.73$.
These colors are typical of a population dominated 
by F5--F8 main-sequence stars, suggesting an old age. 
Assuming a single burst of star formation, 
we infer (using Starburst99, Leitherer et al.~1999) 
a mass of $\approx 10^6 M_{\odot}$ for the galaxy 
and an age $\ga 10^9$ yr for the dominant 
component of its stellar population.

On top of this old component, the galaxy shows 
two bright clusters and a few more, much fainter lumps.
The two brightest clusters have luminosities $M_B \approx -7.2$ 
and $M_B \approx -7.1$, and colors consistent 
with an age $\sim 10^7$ yr and masses of $\sim$ a few $\times 
10^3 M_{\odot}$. Their brightness and morphology 
is consistent with the bright star-forming complexes often found 
in dIrr galaxies (Parodi \& Binggeli 2003).
We obtain that the percentage of flux in the $B$ band 
due to these star-forming complexes (``lumpiness index'') 
is $\approx 7\%$ of the total B-band flux, 
the same value found for a large 
sample of irregulars and spirals regardless of Hubble type 
and galactic mass (Elmegreen \& Salzer 1999; Parodi \& Binggeli 2003).
Thus, the lumpiness index is thought to be a measure 
of star-formation efficiency. These considerations 
support the idea that the bright lumps are indeed 
clusters in the dIrr satellite galaxy and not 
simply background or foreground stars in NGC\,4559.
It is possible that this later episode of star formation 
in the dIrr satellite may have been triggered as this small galaxy 
passed through the disk of NGC\,4559, shocking 
its gas and creating the expanding star-forming wave. 
The non-spherical appearance 
of the large star-forming complex in NGC\,4559 may be due to 
an oblique impact, probably from the south-east to the 
north-west direction (because the oldest 
stars in the field are found in the south-east 
sector, with ages $\ga 20$ Myr; Soria et al.~2004).
The relative masses of the ``bullet'' and swept-up gas 
are also consistent with the results of Comer\'{o}n \& Torra (1994) 
for the case of Gould's Belt. 
The projected distance of the dIrr from the centre 
of the star-forming ring or bubble is $\approx 400$ pc, corresponding 
to a projected relative velocity of $\approx 30$ km s$^{-1}$ 
over 15 Myr.

Thus, we could view the star-forming complex in NGC\,4559 
as a small-scale version of the Cartwheel galaxy, 
where many young ULXs have been detected in the expanding, 
star-forming ring (Gao et al.~2003).
Apart from the different time and length scales involved, 
the main difference between the two systems is that, 
in the Cartwheel, the initial perturbation causing 
the expanding density wave is due to the gravitational 
interaction between the two galaxies; in the case of NGC\,4559, it is
more likely due to the hydrodynamical interaction between 
the gas in the satellite galaxy and the gas-rich disk.


\section{Environmental conditions favourable to ULX formation}
\label{sec:environs}

NGC\,4559 X7 offers an example of a bright ($L_{\rm x} > 10^{40}$) 
ULX in a low-metallicity environment disturbed by close galaxy 
interactions. At least one or both of these elements 
seem to be a common feature for many of the galaxies 
hosting ULXs (eg, galactic interactions for the Antennae, 
the Cartwheel, the M81/M82 galaxy group; 
low metal abundance for the Cartwheel ring, 
the M81 group dwarfs, IZw18). A connection between ULX formation 
and low-Z environment was already suggested 
in Pakull \& Mirioni (2002).

Assuming that most ULXs can be explained by accreting 
black holes more massive ($\approx 50$--$100 M_{\odot}$) 
than those found in nearby X-ray binaries, 
we speculate that these two environmental conditions 
may be most favourable for producing massive remnants:
\begin{itemize}
\item galaxy mergers and close interactions, 
and collisions with satellite galaxies and 
high-velocity HI clouds favour clustered star formation. 
The core of young star clusters may be an environment 
where massive remnants are formed (through the Spitzer 
instability, runaway core collapse 
and merger of the O stars; see Portegies Zwart \& McMillan 2002; 
Rasio, Freitag, \& G\"{u}rkan~2003).
One of the open questions is what type of cluster (ie, what mass range) 
offers the best chance for the core collapse/stellar coalescence 
process to occur within the lifetime of its O stars: 
superstar clusters ($M \ga 10^5 M_{\odot}$) or smaller clusters 
($10^3 < M \la 10^4 M_{\odot}$)?
\item low metal abundance implies a lower mass-loss rate 
in a wind ($\dot{M}_{\rm w} \sim Z^{0.6}$) for the black hole 
progenitor, leading to a larger core and a more massive remnant.
Metallicity also affects the evolution of the donor star  
(for example, metal-poor stars spend a longer fraction of 
their life as red supergiants) and the orbital separation 
of the binary components (which increases 
for a higher mass-loss rate in a wind, 
that is, for a higher non-conservative mass transfer).
This may affect the timescale in which 
the donor star fills the Roche lobe (that is, the timescale 
for the ULX phase).
\end{itemize}

\acknowledgements

Thanks to Helmut Jerjen, Craig Markwardt, Christian Motch, 
Richard Mushotzky and Kinwah Wu, 
who gave important contributions to the work we have presented 
at this conference.

\end{document}